\documentclass[12pt]{article}
\usepackage{amsmath, amsthm}
\usepackage{graphicx}
\usepackage{enumerate}
\usepackage{float}
\usepackage{natbib}
\usepackage{color}
\usepackage{xcolor}
\usepackage{hyperref}
\usepackage{url} 
\usepackage{tikz}
\usetikzlibrary{arrows.meta, positioning, shapes, fit, calc}
\counterwithin{figure}{section}
\counterwithin{table}{section}

\newcommand{\blind}{1}

\usepackage[doublespacing]{setspace}

\begin{document}

\def\spacingset#1{\renewcommand{\baselinestretch}%
{#1}\small\normalsize} \spacingset{1}

\if1\blind
{
  \title{\bf Exploration, Confirmation, and Replication in the Same Observational Study: A Two Team Cross-Screening Approach to Studying the Effect of Unwanted Pregnancy on Mothers' Later Life Outcomes}
  \author{Samrat Roy\\
    Operations and Decision Sciences, Indian Institute of Management Ahmedabad\\
    Marina Bogomolov \\
    Data and Decision Sciences, Technion - Israel Institute of Technology\\
    Ruth Heller \\
    Department of Statistics and Operations Research, Tel-Aviv University\\
    Amy M. Claridge \\
    Child Development and Family Science, Central Washington University\\
    Tishra Beeson\\
    Department of Health Sciences, Central Washington University\\
    and \\
    Dylan S. Small\thanks{This work was supported by Israeli Science Foundation under Grant 406/24.}\hspace{.2cm} \\
    Department of Statistics and Data Science, University of Pennsylvania\\}
  \maketitle
} \fi

\if0\blind
{
  \bigskip
  \bigskip
  \bigskip
  \begin{center}
    {\LARGE\bf Exploration, Confirmation, and Replication in the Same Observational Study: A Two Team Cross-Screening Approach to Studying the Effect of Unwanted Pregnancy on Mothers' Later Life Outcomes
    }
\end{center}
  \medskip
} \fi

\bigskip
\begin{abstract}
The long-term consequences of unwanted pregnancies carried to term on mothers have not been explored much. We use data from the Wisconsin Longitudinal Study (WLS) and propose a novel approach, namely two team cross-screening, to study the possible effects of unwanted pregnancies carried to term on various aspects of mothers' later life mental health, physical health, economic well-being, and life satisfaction. Our approach, unlike existing approaches to observational studies, enables investigators to perform exploratory data analysis, confirmatory data analysis, and replication in the same study. This is a valuable property when there is only one data set available with unique strengths. In two team cross-screening, the investigators split themselves into two teams and the data is split as well according to a meaningful covariate. Each team then performs an exploratory data analysis on its part of the data to design an analysis plan for the other part of the data. The complete freedom of the teams in designing the analysis has the potential to generate new unanticipated hypotheses in addition to a prefixed set of hypotheses. Moreover, only the hypotheses that looked promising in the data each team explored are forwarded for analysis (thus alleviating the multiple testing problem). These advantages are demonstrated in our study of the effects of unwanted pregnancies on mothers' later life outcomes.
\end{abstract}

\noindent%
{\it Keywords: Causal Inference, Multiple Testing, Sample Splitting, Replicability.}  
\vfill

\newpage
\spacingset{1.9} 
\section{Introduction}
\label{intro}

\subsection{An Observational Study Assessing the Effect of Unwanted Pregnancies on Mothers’ Later Life Outcomes}
\label{intro_1}
Unwanted pregnancies are prevalent worldwide \citep{bearak2018global}.
For unwanted pregnancies carried to term, it is crucial to assess their effect on the child who is born and the mother to design policies that best support the child and the mother. For the mother, there is evidence of short-term effects such as less emotional attachment to her baby during pregnancy, greater parental stress at 6 months and one year postpartum, and less effective parenting strategies \citep{pakseresht2018physical}. However, the long term effects have not been studied much. The only study we are aware of is \cite{herd2016implications} that used linear regression to examine one aspect of mental health, depression. In this work, our aim is to study the possible effects of an unwanted pregnancy carried to term on various aspects of a mother's later life mental health, physical health, economic well-being, and life satisfaction. A data set that has unique strengths to carry out such an observational study is the Wisconsin Longitudinal Study (WLS) \citep{herd2014cohort}. The WLS randomly sampled one-third of people graduating from high school in Wisconsin in 1957 and has followed them longitudinally every decade since. Strengths of the WLS include:
\begin{itemize}
    \item[\textbf{--}]The WLS asked if the pregnancy was wanted somewhat contemporaneously at about 36 years of age. Women who had experienced pregnancies by the age of 36 were asked for each pregnancy: `Before you became pregnant, did you want to become pregnant at that time?' and if the woman answered no, `Did you want to have another baby sometime?". If the woman answered no to both questions, it was coded as an unwanted pregnancy. This way of identifying an unwanted pregnancy when the woman is of age 36 is better than asking a woman when a child is completely grown up, but not as good as asking a woman about her pregnancy intention prior to her pregnancy, as her recall could be colored by the events of the intervening years between the pregnancy and the time the woman turned 36. However, we proceed with this way of identification due to the lack of any other reliable data sources. 
    \item[\textbf{--}]Prospectively before the birth of the participants' children, the WLS measured a wide range of variables that could confound the relationship between unwanted pregnancies and later life outcomes. These variables include family history, adolescent characteristics, educational and occupational achievements, and aspirations.
    \item[\textbf{--}]The WLS has long-term longitudinal follow-up of its respondents when they were around age 53, 65 and 72, allowing the study of later life outcomes. 
\end{itemize}
Note that women in the WLS experienced most of their pregnancies before the 1973 \textit{Roe v. Wade} Supreme Court decision that made abortion legal throughout the United States. Before 1973, abortion was illegal in Wisconsin. Thus, this dataset is most relevant to the question of what is the effect of unwanted pregnancy in places where abortion is illegal. This is an important current question because abortion is currently completely illegal in 24 countries, legal only when the women’s health is at risk in 37 other countries, and illegal in some U.S. states following the 2022 {\it{Dobbs v. Jackson}} decision.  
\subsection{Exploratory Data Analysis, Confirmatory Data Analysis and Replication in Observational Studies} 
\label{intro_2}
Exploratory data analysis (EDA), confirmatory data analysis, and replication are three important aspects of building strong evidence from observational studies. EDA helps form hypotheses that might not have been anticipated and remove outliers that do not represent natural variation in the data 
\citep{tukey1977exploratory}.
It can also help identify a variable that does not measure what one thought it did and may suggest a different variable that better measures the construct that one is actually interested in. 

John Tukey coined the term EDA and developed many foundational techniques for it, but he also stressed the importance of confirmatory data analysis.  Likening EDA to detective work about a crime and confirmatory data analysis to a scientific trial of the hypotheses suggested by EDA \citep{tukey1977exploratory}, Tukey said \citep{tukey1980we}: 
\textit{``Important questions can demand the most careful planning for confirmatory analysis."}   

Replication of results is also important for science. \textcolor{blue}{\cite{susser1973causal} emphasized the importance of replication in observational studies, defining replication as consistency with respect to different ``times, circumstances and people." Replicating an observational study is particularly helpful when the concerns about hidden bias in the original and replicated studies are different \citep{rosenbaum2001replicating}. One way this can happen is when the reasons for receiving treatment in the two studies differ \citep{rosenbaum2001replicating, rosenbaum2015see}.} For example, if a study in New York City found that fish eaters live longer than non-fish eaters, then repeating the study in Philadelphia would provide limited additional information as both studies could be biased because fish is a comparatively expensive food in cities, favored by those seeking to maintain a healthy diet \citep{rosenbaum2015see}; a more helpful replication would be the study where \cite{lund1993reduced} compared wives of fishermen in Norway to wives of other workers of similar income. The reasons why some subjects in the New York City study eat more fish than others (e.g., the fish eaters have higher income and are more health conscious) differ from the reasons why some subjects in the Norway study eat more fish than others (e.g., the wives of fishermen eat fish that was caught by their husband but could not be sold on the market).  If both studies found evidence of a treatment effect, then in order to just explain this away as bias without any treatment effect, one would need to posit that two different biases are present {\textendash} both a bias that those who eat more fish in New York City are doing other things which improve their health compared to those who don't eat much fish and a bias that wives of fishermen in Norway are doing other things than eating a lot of fish which improved their health compared to  wives of other workers of similar income. However, if one had studied New York City and Philadelphia, an ostensible treatment effect in both studies could be explained away by just one bias that those who eat more fish in big cities are doing other things which improve their health compared to those who don't eat much fish.  

Typically, EDA, confirmatory data analysis, and replication are performed sequentially using more than one dataset. However, sometimes there is a single dataset with unique strengths (e.g., the WLS to study the long-term effects of unwanted pregnancy on mothers) and it is desirable to be able to perform all of EDA, confimatory data analysis and replication using this one dataset.  In this paper, we develop an approach called {\it{two team cross-screening}} for performing EDA, confirmatory data analysis, and replication using a single observational study dataset.

\vspace{0.1in}
The rest of the paper is organized as follows. \S~\ref{intro_3} describes our two team cross-screening approach, and \S~\ref{intro_4} summarizes the existing relevant literature. 
\S~\ref{trt_out} describes our application of two team cross-screening to study the effects of unwanted pregnancy. \S~\ref{auto_results} compares the results of our approach to those of other existing approaches. \S~\ref{esize} discusses using two team cross-screening to obtain confidence intervals for effect sizes, which is followed by a discussion in \S~\ref{disc}.

\section{Exploration, Confirmation and Replication in the Same Observational Study: Two Team Cross-Screening}
\label{intro_3}
Our proposed two team cross-screening approach primarily consists of the following steps:

\textbf{Step 1 - \textcolor{blue}{Pre-Analysis Discussion}:} 
This step involves a discussion among all researchers that is held before any analysis is performed. In this step, researchers are allowed to see the names and descriptions of the variables that were measured in the study but are only partially allowed to view the data. Specifically, they can explore the joint distribution of the covariates and the treatment, denoted by $(X_i)_{i\in [n]}$ and $(T_i)_{i\in [n]}$ respectively, where $n$ is the number of subjects in the study \citep{rubin2007design, small2024protocols}. In this step, the researchers also discuss how the data will be split. As discussed in \S~\ref{intro_2}, the splitting factor could be just a time point or location, but a particularly helpful splitting factor for addressing concerns about unmeasured confounding is two subgroups that receive treatment for relatively different reasons.  
In this step, researchers also identify the outcomes of interest and covariates to be controlled for. Discussing potentially relevant outcomes increases the chance of replicability, since achieving replicability for a given outcome requires that both teams include that outcome in their analysis plans. After completing the pre-analysis discussion, the researchers divide themselves into two teams, A and B, each ideally comprising both statistician(s) and domain expert(s). Each team is assigned one part of the data, corresponding to one level of the splitting factor, and will have no access to the other part \footnote{Technically, as discussed in Step 1, the team could view the treatment and covariates in the other part of the data}.

\textbf{Step 2 - \textcolor{blue}{Exploratory Data Analysis (exploration)}:} In this step, each team performs an exploratory data analysis on its assigned portion of the data without access to the other part. As noted in the previous step, both teams examine the set of key outcomes determined during the pre-analysis discussion. In addition to this predefined set of outcomes, the teams are also permitted to investigate new unanticipated hypotheses. Given a target level $\alpha$ for the family-wise error rate (FWER) of the entire analysis, Team A explores its portion of the data and, based on its findings, constructs an $\alpha/2$ (e.g., 0.025 for $\alpha=0.05$) FWER testing plan for the other part of the data. Likewise, Team B explores its portion and develops an $\alpha/2$ FWER testing plan for the other part of the data. This exploratory stage helps the teams design efficient tests by allocating their $\alpha/2$ judiciously and allows the generation of new, unanticipated hypotheses alongside the initially planned ones.  

\textbf{Step 3 - \textcolor{blue}{Confirmatory Data Analysis (confirmation)}:} Team A shares its prepared $\alpha/2$ FWER testing plan along with the code needed to implement it, which is then executed on Team B’s portion of the data. Similarly, Team B shares its $\alpha/2$ FWER testing plan and the accompanying code, which is run on Team A’s portion of the data.

\textbf{Step 4 - \textcolor{blue}{Reporting Global Null and Replicable Findings}}: Once the testing plans proposed by both teams are executed on the other team’s portion of the data, if a hypothesis is rejected by either team, then the global null for that hypothesis is rejected (e.g., if the null hypothesis of no treatment effect on an outcome is rejected for one part of the data, then the global null of no treatment effect for the whole data is rejected).  The overall FWER for global null tests is controlled at $\alpha$ by the Bonferroni inequality \citep{cross-screening}. Any hypothesis rejected under the plans of both teams is considered a replicable finding and the probability of making at least one false replicability claim is controlled at $\alpha$ \citep{repl_biometrika}. 

Figure \ref{plan_summary} summarizes the steps in the proposed two team cross-screening approach.

\clearpage
\begin{figure}[h]
\label{plan_summary}
\centering
\resizebox{0.9\textwidth}{!}{%
\begin{tikzpicture}[
  node distance=1.6cm and 2.8cm,
  every node/.style={font=\Large, align=center}, 
  action/.style={draw=black, thick, rectangle, rounded corners,
                 fill=gray!10, minimum width=3.2cm, minimum height=1cm},
  data/.style={draw, rectangle, rounded corners, fill=white,
               minimum width=3.2cm, minimum height=0.8cm},
  comp/.style={draw=red, thick, rectangle, rounded corners,
               fill=gray!10, minimum width=3.2cm, minimum height=1cm},
  arrow/.style={-{Latex[length=3mm]}, ultra thick}
]

\node[data] (characteristics) {Data characteristics\\(names and description of variables)};

\node[action, below=of characteristics] (discussion)
  {Pre-analysis discussion (by all researchers)};

\node[data, left=3.8cm of discussion] (XE) {$(X_i, T_i)_{i\in [n]}$};

\draw[arrow] (characteristics.south) -- (discussion.north);
\draw[arrow, dashed] (XE.east) -- (discussion.west);

\node[data, below=1.6cm of discussion] (codespecs)
{List of outcomes of interest and covariates to control for; \\
Decision on how to split the data and researchers into two parts};

\draw[arrow] (discussion.south) -- (codespecs.north);

\node[data, right=1.8cm of codespecs] (XEY) {$(X_i, T_i, Y_i)_{i\in [n]}$};

\node[comp, below=1.6cm of codespecs] (codesplitcomp)
{Splitting the data into two parts with indices  \\  $\mathcal I_1,  \mathcal I_2$: $\mathcal I_1\cap \mathcal I_2 = \emptyset, \mathcal I_1\cup \mathcal I_2 = [n]$};

\draw[arrow] (codespecs.south) -- (codesplitcomp.north);      

\draw[arrow] (XEY.south west) to[out=-90, in=45] (codesplitcomp.north east);

\node[data, below left=1.4cm and 1.4cm of codesplitcomp] (part1)
  {Part 1: $(X_i, T_i, Y_i)_{i\in \mathcal{I}_1}$};
\node[data, below right=1.4cm and 1.4cm of codesplitcomp] (part2)
  {Part 2: $(X_i, T_i, Y_i)_{i\in \mathcal{I}_2}$};

\draw[arrow] (codesplitcomp.south west) to[out=-120, in=90] (part1.north);
\draw[arrow] (codesplitcomp.south east) to[out=-60, in=90] (part2.north);

\node[action, below=of part1] (EDA_A) {EDA by Team A};
\node[action, below=of part2] (EDA_B) {EDA by Team B};

\draw[arrow] (part1) -- (EDA_A);
\draw[arrow] (part2) -- (EDA_B);

\node[data, below=of EDA_A] (codeA)
  {Team A's confirmatory design };
\node[data, below=of EDA_B] (codeB)
  {Team B’s confirmatory design};

\draw[arrow] (EDA_A) -- (codeA);
\draw[arrow] (EDA_B) -- (codeB);

\node[comp, below=of codeA] (confirmA)
  {Confirmatory data analysis };
\node[comp, below=of codeB] (confirmB)
  {Confirmatory data analysis };

\draw[arrow] (codeA) -- (confirmA);
\draw[arrow] (codeB) -- (confirmB);

\draw[arrow] (part1.south east) to[out=-45, in=135, looseness=1.2]
  (confirmB.north west);
\draw[arrow] (part2.south west) to[out=-135, in=45, looseness=1.2]
  (confirmA.north east);

\node[data, below=of confirmA] (R2) {Indices of rejected hypotheses: $R_2$};
\node[data, below=of confirmB] (R1) {Indices of rejected hypotheses: $R_1$};

\node[action, below=2cm of $(R1)!0.5!(R2)$] (reporting)
  {Reporting replicable findings: $R_1\cap R_2$\\
   Reporting global null findings: $R_1\cup R_2$};

\draw[arrow] (R2) to[out=-60, in=150] (reporting);
\draw[arrow] (R1) to[out=-120, in=30] (reporting);
\draw[arrow] (confirmA) -- (R2);
\draw[arrow] (confirmB) -- (R1);

\end{tikzpicture}%
}
\caption{Schematic of the two team cross-screening procedure. Gray rectangles denote actions with black borders denoting researcher actions and red borders denoting computer-executed actions. Solid arrows indicate inputs and outputs. $(X_i, T_i, Y_i)_{i \in [n]}$ represent the full data for $n$ subjects in the study, where $X_i$ are the covariates, $T_i$ is the treatment and $Y_i$ are the outcomes for subject $i$. The dashed arrow from $(X_i, T_i)_{i \in [n]}$ to pre-analysis discussion indicates that the input is optional (see \S~\ref{intro_3} for details).}
\label{plan_summary}
\end{figure}

\section{Literature Review}
\label{intro_4}
There has been work that combines some but not all of exploration, confirmation, and replication in the same observational study.

\noindent \textbf{EDA but no replicability}: In a single-split sample design \citep{heller2009split}, the data is randomly split into a smaller planning sample and a larger analysis sample, where the planning sample is used to plan the analysis. EDA can be conducted on the planning sample, and then the analysis sample is used for confirmatory analysis. \textcolor{blue}{However, this approach does not provide an opportunity for replication across the planning and analysis samples because hypotheses are tested only on the analysis sample.} Another related work is \cite{yu2011bayesian} in which  multiple data analysts independently examine randomly selected portions of a dataset and perform subjective Bayesian analyses, and the results are aggregated using a Bayesian synthesis method. This approach does not explicitly address 
replicability.
\vspace{0.1in}\\
\noindent \textbf{Replicability but no EDA}: One team automated cross-screening \citep{repl_biometrika, cross-screening} is an approach involving one team in which the team divides the data into two parts, I and II. Given a target level $\alpha =0.05$ for the FWER of the entire analysis, it prespecifies, without exploring the data, automatic rules for using  part I to define  a $0.025$ FWER controlling procedure  for  part II and vice versa. 
For example, based on part I, one might test 
all prespecified outcomes, select those with $p$-value less than $0.025$ in part I of the data, and test them in part II  using the  \cite{holm1979simple} procedure at level $0.025$. For valid FWER control, this automated cross-screening approach requires that the way part II of the data will be analyzed to choose an analysis for part I be automatic (i.e., specified before looking at the data).  Automated cross-screening cannot accommodate exploring both parts of the data. The reason is that if the researchers used EDA in part I of the data to choose hypotheses on part II of the data and then went to do EDA in part II of the data to choose hypotheses on part I of the data, the EDA in part II would be biased because they already knew what the EDA in part I of the data yielded. This bias may result in an inflated FWER for the global null hypotheses as well as for replicability findings. Having two independent teams prevents this bias, allowing unbiased exploration of both parts of the data and FWER control. 

Another approach that allows for replicability but not EDA is evidence factors \citep{rosenbaum2015see, rosenbaum2017general, karmakar2019integrating}, which are statistically independent tests of a treatment effect in the same observational study where the potential biases of the tests are different. The tests can be based on two separate pieces of the data or overlapping aspects of the data. If the tests concur in rejecting a null hypothesis of no treatment effect after accounting for multiplicity, then the finding is replicated. The tests are specified in advance of examining the data, so evidence factors allow for replicability but not EDA.  

Unlike automated cross-screening and evidence factors, two team cross-screening allows investigators to perform EDA prior to making the analysis plan, which can help in producing efficient designs and generate new unanticipated hypotheses in addition to the set of hypotheses that the investigators initially planned to test.     

\section{Two Team Analysis of Unwanted Pregnancy Data}
\label{trt_out}
\subsection{Pre-Analysis Discussion}
\label{pre-discussion}
We first discussed what variable to split on. As described in \S~\ref{intro_3}, to assess replicability with respect to different sources of hidden bias, the data should be split based on a factor that individuals with one level of the factor could receive treatment for relatively different reasons than individuals with another level of that factor. In our study of the effect of unwanted pregnancy on mothers, we split on Catholic vs. non-Catholic women.  Some Catholic women could have unwanted pregnancies because they followed the Catholic Church’s opposition to contraceptive methods \citep{miller2002religiousness, jones2011countering} [note that it would be good if we could verify that the Catholic Church’s opposition actually had an effect on contraception among Catholic women, but unfortunately there are no data in the WLS to verify this]. On the other hand, the reason for non-Catholic women to have an unwanted pregnancy could be relatively more likely to be because a woman (or her partner) either willingly did not use contraception or improperly used contraception. If we were to find that unwanted pregnancy was associated with increased depression among both Catholic and non-Catholic women, this could not be explained away as non-causal just by Catholic women who followed the Catholic Church's opposition to contraceptive methods having a more depressive personality than Catholic women who ignored the Church's opposition or just by non-Catholic women who improperly used contraception having a more depressive personality than non-Catholic women who properly used contraception.  Both biases (or some other biases) would have to be present to explain away the associations as non-causal.

\begin{figure}
\centering
\includegraphics[scale=0.8]{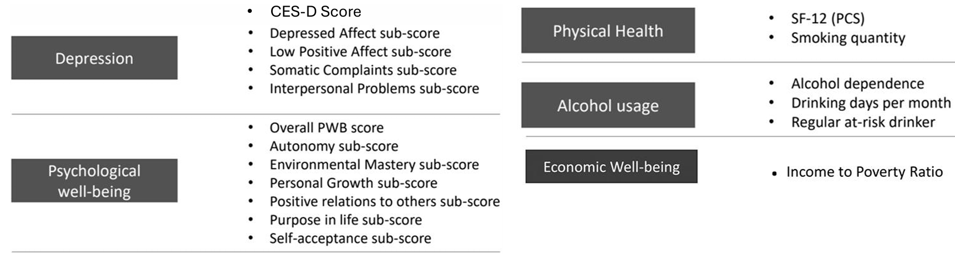}
\caption{List of outcomes decided during pre-analysis discussion. The CES-D score is an overall measure of depression that aggregates the four subscale scores detailed below it \citep{orme1986factorial}.}\label{fig:outcomes}
\end{figure}

After deciding how to split the data, we discussed outcomes we might want to consider. As explained in \S~\ref{intro_3}, this discussion improves the chances of replicability by increasing the chance that both teams would consider at least some of the same outcomes for inclusion in their analysis plans. We decided to focus on five aspects of later life outcomes shown in Figure \ref{fig:outcomes}, each comprising several measures summarized there.  
See details in Supplementary Material-1 (SM-1),  \S~A. As discussed in \S~\ref{intro_3} and \S~\ref{plans}, in Step 2 (EDA), the teams will explore the outcomes in Figure \ref{fig:outcomes} but their EDA might also generate some new outcomes of interest (the WLS contains thousands of outcomes). 

Next we discussed how to control for confounding and what potential confounders to control for. To control for confounding, we decided to use risk-set matching \citep{lu2023risk} for which we now explain the rationale. In an observational study, a direct comparison of treated to control subjects may be biased because of confounders, i.e., pre-treatment differences between the treatment and control groups that affect the outcome. One way to reduce potential bias from potential confounders is to match treated subjects to control subjects on observed potential confounders, then compare treated to control subjects within these matched sets and then aggregate these comparisons \citep{rosenbaum2020modern}. If all potential confounders have been matched on, then the matched observational study can be analyzed like a stratified randomized experiment \citep{rosenbaum2020modern}. In standard matching, treatment occurs at a single time, and treated subjects are matched to controls  on all observed potential confounders at that time. However, in our study, the treatment (an unwanted pregnancy) can occur at different times in a woman’s life. When treatment occurs at different times and treatment can affect future values of confounding variables, standard matching can be biased. Suppose we want to match on education and use usual matching to match a woman who had an unwanted pregnancy at age 18 to a woman who did not have an unwanted pregnancy. The woman who had an unwanted pregnancy at the age of 18 might not have gone to college because of the unwanted pregnancy, so her education is not comparable to a woman who did not have an unwanted pregnancy. Here, education is partially a post-treatment variable, and matching on it can cause bias \citep{rosenbaum1984consequences}. Risk-set matching avoids such bias by matching sequentially over time only on potential confounders that have been observed up to the time of treatment. In risk-set matching, we would match a woman who became pregnant at age 18 to a woman who had not yet become pregnant by age 18 on education up to age 18, the time of pregnancy. In addition to education up to the time of the pregnancy, we decided to match for variables that measure a woman's sociopsychological characteristics up to the time of the pregnancy and her childhood socioeconomic status. \S~A of SM-1 provides details.  Risk-set matching that produced 325 matched pairs for the Catholics and 383 matched pairs for the non-Catholics. Figure D.2 in SM-1 depicts the Love plots of the absolute values of the pre-matching and post-matching standardized differences (that is, average difference between the treated units and matched controls in standard deviation units) for both the Catholic and non-Catholic subgroups. The absolute values of the post-matching standardized differences were all less than 0.2, which indicates acceptable balance of the covariates between the treated and matched control groups \citep{silber2001multivariate}. 

Finally, we divided ourselves into two independent teams, A and B. Each team consisted of two statisticians and one maternal health researcher. Team A was assigned the Catholic data, and Team B was assigned the non-Catholic data. 

\subsection{Exploratory Data Analysis}
\label{plans}
Team A analyzed the Catholic data to plan the analysis for the non-Catholic data. Team B analyzed the non-Catholic data to plan the analysis for the Catholic data. To maintain the overall FWER control at $\alpha=0.05$, we split the whole $\alpha$ equally between the two teams, each having an allowance of spending $0.025$ in total. In the following, we provide a detailed description of the analysis plans prepared by both teams.

\subsubsection{Analysis Planned by Team A} 
\label{plan_a}
In this section, we, team A, explored the data on the Catholic subgroup to plan the analysis for the non-Catholic data. We first examined the variables in Figure \ref{fig:outcomes} when the women were approximately 53 years old. As summarized in Table D.1 in SM-1, the depression score turned out to be significantly higher among the treated group (women with an unwanted pregnancy) with p-value  $0.017$ (one-sided Wilcoxon signed-rank test). We also considered the subscales of the depression score in addition to the overall score, but decided based on the p-values and interpretability that it would be best to just focus on the overall score in our analysis plan. For psychological well-being, the self-acceptance score was significantly lower among the treated group (see Table D.2 in SM-1), and we decided to include the self-acceptance score in our analysis plan.  We investigated depression and self-acceptance at ages 65 and 72 and found the same direction of effects as at age 53, but decided, based on p-values and for clarity of focus, just to test depression and self-acceptance at age 53 in our analysis plan. For physical health, we found that the mean SF-12 PCS score for the treated and matched control groups were quite similar and the p-value for the test of a difference between the groups was 0.53 (one-sided Wilcoxon signed-rank test). Hence, we decided not to include the testing of SF-12 score in our plan. For smoking quantity, defined by the mean number of packs of cigarettes the woman usually smokes (or used to smoke) per day, the corresponding p-value was $0.22$, and therefore we did not find any evidence of differential behavior in smoking, and the corresponding test was not included in our plan. For alcohol usage also, none of the three variables, namely, drinking days, regular `at risk' drinker, and possible alcohol dependence (see \S~A of SM-1 for definitions), showed any significant difference between the treated and matched control groups.     

For economic well-being, we initially found a significantly higher income to poverty level ratio (see \S~A of SM-1 for definition) among the treated group (women with an unwanted pregnancy). We also found significantly higher total personal income and total household income (in the last $12$ months prior to the moment when the participants answered the questionnaire) among the treated group. We found this result surprising, if anything, we had expected that having an unwanted pregnancy would decrease income. To better understand potential reasons for this surprising result, we considered the following potential mediators: a) Social participation, that is the level of involvement with social organizations such as church connected groups, lodges, sports teams, neighborhood improvement organizations and so on (number of organizations in which the woman had at least some involvement which ranged from 0 to 14), b) Number of job spells after the index time (i.e., the time at which the treated woman had her first unwanted pregnancy), which is a measure of the job stability of the woman, c) Number of divorces or separations after the index time, d) Additional years of education after the index time and e) Additional number of children from further unwanted pregnancies after the index unwanted pregnancy. Among these five variables, we found that unwanted pregnancy significantly (i) decreased job stability (that is, more job spells); (ii) increased additional children from further unwanted pregnancy beyond the index unwanted pregnancy and (iii) increased number of divorces or separations.   

Examining these potential mediators produced interesting findings, but did not explain why unwanted pregnancy would lead to higher income. We could see how decreased job stability, increased additional children from further unwanted pregnancies, and increased divorces or separations could lower income, but not how they could increase income. We decided to further investigate the different components of income which revealed that the two key components, namely (i) wages and (ii) earnings from own business, had no significant difference between the two groups. However, the treated group with unwanted pregnancies had significantly higher pensions, annuities, and survivor’s benefits compared to the controls. This difference in income only due to pensions, annuities, and survivor’s benefits was not interpretable to us and we decided not to focus on it. Thus, although we initially found the overall income difference was statistically significant, a more careful data exploration revealed that the overall income difference was not actually meaningful to us. This was an instance where EDA prevented us from spending $\alpha$ on a hypothesis that was not worthwhile. However, our investigations of possible mediators of income difference led us to additional hypotheses of interest about unwanted pregnancy's effects that we did not plan initially. This is an advantage of looking deeper into the data that allows one’s mind to wander creatively \citep{maceachern2019preregistration}.

In addition to exploring different hypotheses, cross-screening gives us a chance to select different test statistics for the chosen hypotheses to achieve better power. To that end, as used in \cite{cross-screening}, we explored the family of U-statistics introduced in \cite{rosenbaum2011new} and reviewed in \S~B of the SM-1. Specifically, we explored the U-statistic with three sets of values for $m$, $\underline{m}$, and $\overline{m}$ as $(8,5,8)$, $(8,6,7)$ and $(8,7,8)$, along with the Wilcoxon signed-rank test. For each of the hypotheses, we selected the test statistic for which the corresponding p-value was minimum (see Table D.6 in SM-1). 

\begin{figure}[h!]
    \centering
    \includegraphics[scale=0.65]{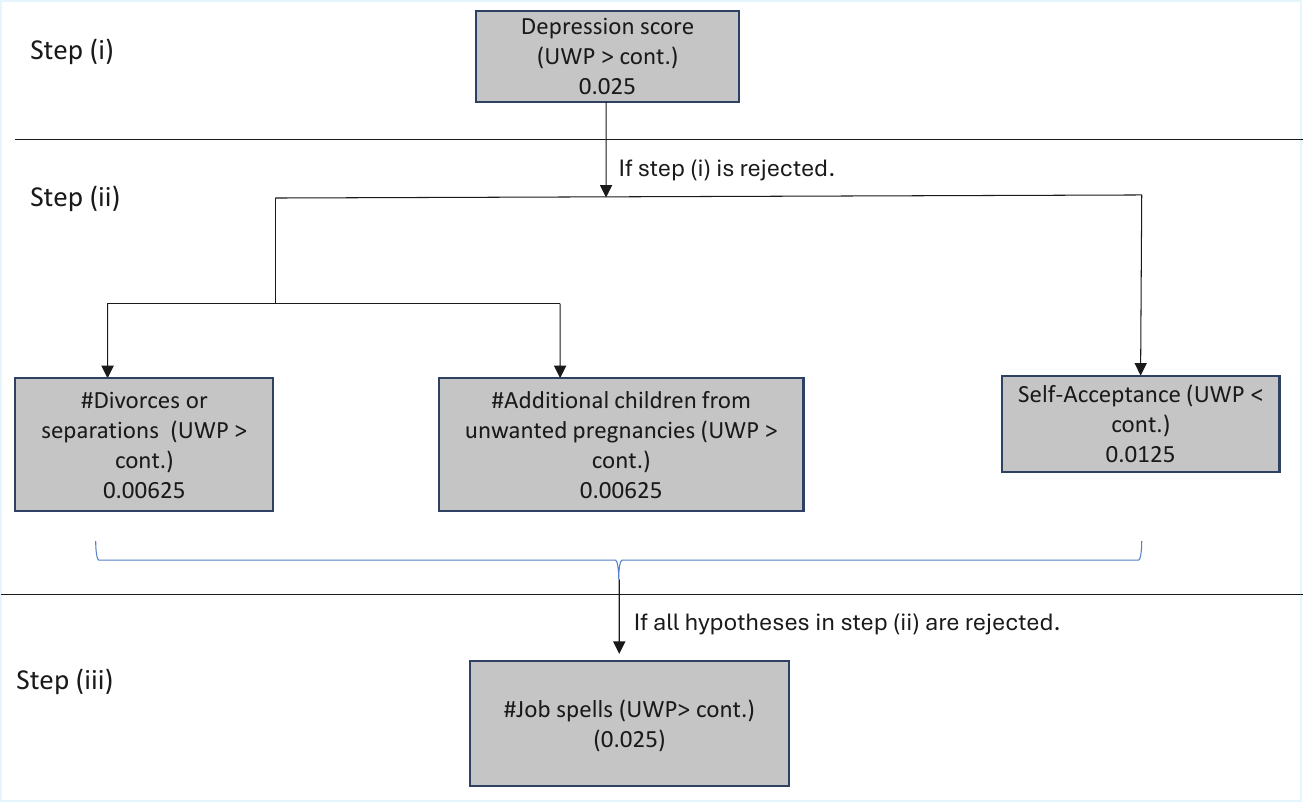}
    \caption{Analysis plan prepared by Team A based on the Catholic data, which will be performed on the non-Catholic data: at each step, the hypothesis (or the family of hypotheses) is depicted along with the direction of the alternative (UWP stands for unwanted pregnancy and cont. for control) and the corresponding amount of $\alpha$ that will be spent for the testing of that hypothesis.} 
    \label{fig:test_seq}
\end{figure}

To control for the multiple testing of the five hypotheses listed in Table D.6 and put priority on the most important hypotheses, we followed the idea of serial gatekeeping procedures in which the hypotheses were divided into families and the priority families served as `gatekeepers' for subsequent families in the sense that all hypotheses in the current family must be rejected to start testing the hypotheses in the next family \citep{dmitrienko2007gatekeeping}. After considering the importance of different hypotheses and their chances of leading to significant findings, we decided on the following analysis plan, which is shown as a flow chart in Figure \ref{fig:test_seq}.

\begin{itemize}
    \item [(i)] Start with testing whether the depression score is higher among women giving births from unwanted pregnancies, using U-statistic with $(m, \underline{m}, \overline{m})$ as $(8,6,7)$ with $\alpha=$ $0.025$ (see step (i) of Figure \ref{fig:test_seq}). We chose to put the highest priority on depression because of its clinical importance.
    \item [(ii)] If the above test in step (i) is rejected, carry the whole $\alpha=0.025$ forward to the next step and perform the following multiple testing procedure. In the first path, use the first half of $\alpha$, that is, 0.0125, to test whether the self-acceptance subscale score is significantly lower in women with unwanted pregnancies using U-statistics with $(m, \underline{m}, \overline{m})$ as $(8,5,8)$.  In the second path, further split the corresponding $\alpha$, that is $0.0125$, into two equal parts and test in parallel the following hypotheses with the Wilcoxon signed-rank test: (1) whether the women with unwanted pregnancy have significantly higher number of additional children from further unwanted pregnancies after the index time, and (2) whether the women with unwanted pregnancies have significantly higher number of divorces or separations after the index time.

    \item [(iii)] If all three hypotheses in step (ii) are rejected, then proceed further to test whether the number of job spells is significantly higher among mothers who gave birth to unwanted pregnancies with $\alpha=0.025$ (see step (iii) in Figure \ref{fig:test_seq}). For this test, use the U-statistic with $(m, \underline{m}, \overline{m})$ as $(8,7,8)$.    
\end{itemize}

\subsubsection{Analysis Planned by Team B}
\label{plan_b}
In this section, we, team B, analyzed the non-Catholic data, in order to design the analysis for the Catholic data. Our first step was to examine the outcomes depicted in Figure \ref{fig:outcomes}. We used one-sided tests for each outcome assuming no hidden bias (that is, sensitivity parameter $\Gamma=1$ in \cite{rosenbaum2007sensitivity}), as well as under the arguably more reasonable assumption that there was bias due to nonrandom treatment assignment up to $\Gamma=1.2$. The sensitivity parameter $\Gamma$ quantifies the maximum ratio of the odds of having an unwanted pregnancy between two women with similar observed covariates at the time when one of them experienced an unwanted pregnancy \citep{rosenbaum2007sensitivity}. Formally, for women $i$ and $j$ with the same observed covariates, the relationship between probabilities of receiving treatment, $\pi_i$ and $\pi_j$ for women $i$ and $j$ respectively, are bounded by $\frac{1}{\Gamma} \leq \frac{\pi_{i} (1 - \pi_{j})}{\pi_{j} (1 - \pi_{i})} \leq \Gamma$. The sensitivity parameter $\Gamma=1$ means that there is no unmeasured confounding and $\Gamma=1.2$ means that because of an unmeasured confounder, the odds of having an unwanted pregnancy could increase by 20\% among women in the same matched pair. We decided to incorporate $\Gamma=1.2$ into our proposed analysis in addition to $\Gamma=1$ in order to allow for the possibility of moderate bias from unmeasured confounders. This particular choice of $\Gamma=1.2$ was inspired from relevant literature \citep{rosenbaum2007sensitivity, imbens2015causal, ding2016sensitivity}.

The tests were run using the function ``{\it senmv}" from R package ``{\it sensitivitymv}" available in CRAN (\cite{rosenbaum2007sensitivity}). Specifically, we used the permutation t-test statistic ``t", Wilcoxon’s signed-rank test statistic ``W", as well as a test that ignores absolute paired differences smaller than half the median, ``i". The p-values for one-sided hypotheses in the expected direction of effect for the questionnaires are reported in Table D.3 of SM-1. The smoking quantity, alcohol usage, and economic well-being were also examined, and turned out to be all non-significant with large p-values (omitted for brevity). The only outcomes with a p-value at most 0.025 at $\Gamma = 1.2$ were depression score, the low-positive affect subscale score, and the interpersonal problems subscale score. The low-positive affect subscale score stood out since it had the strongest evidence, by a great margin, over all other outcomes with all the considered tests. The low-positive subscale consisted of questions "I felt as good as other people", "I was happy" and "I enjoyed life" (see Table D.4 in SM-1), and women with an unwanted pregnancy agreed significantly less with these statements than their paired controls among non-Catholic women. Since difference in a subscale score implies a  difference in depression, and since the evidence of this subscale score was by far greater than all other subscale scores as well as the aggregate depression score, the low-positive affect subscale score was the primary outcome we considered testing. 

The exploratory stage allowed us to examine more thoroughly our decision to concentrate on the low-positive affect subscale score. Our next step was to repeat the $p$-value calculations for all the outcomes on important subgroups, as well as using different methods to impute missing data. Specifically, we considered only those matched pairs that, at the index time, were at most 2 years apart in age, or had the exact same number of years of education, or had the exact same number of prior children. Moreover, we considered the answers from survey year 2003 in the WLS database (when women were at age 65) whenever it was missing in the survey year 1992 (when women were at age 53). The qualitative conclusions remained unchanged. The quantitative conclusions were similar to the original analysis, so we finalized our decision to test the null hypothesis of no low-positive subscale score effect using the permutation t-test on all pairs, without imputing missing data. Table D.3 in SM-1 shows that the $p$-value using this test on the non-Catholics is $0.0003$ with $\Gamma=1$ and $0.0082$ with $\Gamma = 1.2$. 

Next, we searched for possible effect modification with the covariates considered for the risk-set matching in Figure D.2 of the SM-1. Our aim was to identify the covariates that are associated with the difference in low-positive affect subscale score between the women who had an unwanted pregnancy and their paired controls. We discovered moderate evidence that the aforementioned difference varied across the age when unwanted pregnancy occurred. 
The greatest effect appeared when unwanted pregnancy occurred when a woman was younger. We tested the null hypothesis that the effect of unwanted pregnancy on the low-positive affect subscale for women aged below their mid-twenties was higher than the effect for women aged above their mid-twenties using a one-sided Wilcoxon rank sum test and the p-value was $0.011$. We included this test in our analysis plan for the Catholics. 

Based on the above exploratory data analysis on non-Catholic data, our suggested analysis plan for the Catholics data was to test the following six hypotheses in order \citep{rosenbaum2008testing}, i.e.,   testing proceeds to the next step only if the previous null is rejected and otherwise, stop and report all rejections up to that point. Each test is carried out at level $0.025$. 

(1) Test the null hypothesis that the low-positive affect subscale score distribution is the same for women that had and did not have an unwanted pregnancy, using a one-sided permutation t-test.

(2) Repeat at $\Gamma=1.2$, using the one-sided permutation t-test.

(3)   Test the null hypothesis of no effect modification by age, using the one-sided Wilcoxon rank sum test.

(4)  Test the null hypothesis that the aggregate depression score distribution is the same for women that had and did not have an unwanted pregnancy, using a one-sided permutation t-test.

(5) Test the null hypothesis in (4) at $\Gamma=1.2$, using the one-sided permutation t-test.

(6) Test the  hypothesis in (3)  at $\Gamma=1.2$, using the one-sided Wilcoxon rank sum test.

Although rejection of the null hypothesis of no effect on low-positive subscale score in (1) and (2) also implies rejection of the null hypothesis of no effect on any aspect of depression, we still suggested performing the tests in items (4) and (5) on aggregate depression because their result may be of interest to investigators that are used to reporting of the effect of depression using the aggregate score. Moreover, we put (6) last, since at $\Gamma = 1.2$, we did not have strong evidence of effect modification by age among the non-Catholics that we examined in our EDA, and so we did not want to waste the $\alpha$ of previous hypotheses on this hypothesis. On the other hand, if the testing of this hypothesis is reached for the Catholics, and it is rejected, then it will be possible to state that there is strong evidence towards effect modification by age in the Catholic subgroup, which would be interesting.  
\subsection{Confirmatory Data Analysis and Assessing Replicability}
\label{results}
In this stage, both teams shared their proposed analysis plans and the necessary codes to implement the plan. Team A's plan was applied to non-Catholic data, and similarly, Team B's plan was applied to Catholic data. 
\begin{figure}[h!]
    \centering
    \includegraphics[scale=0.6]{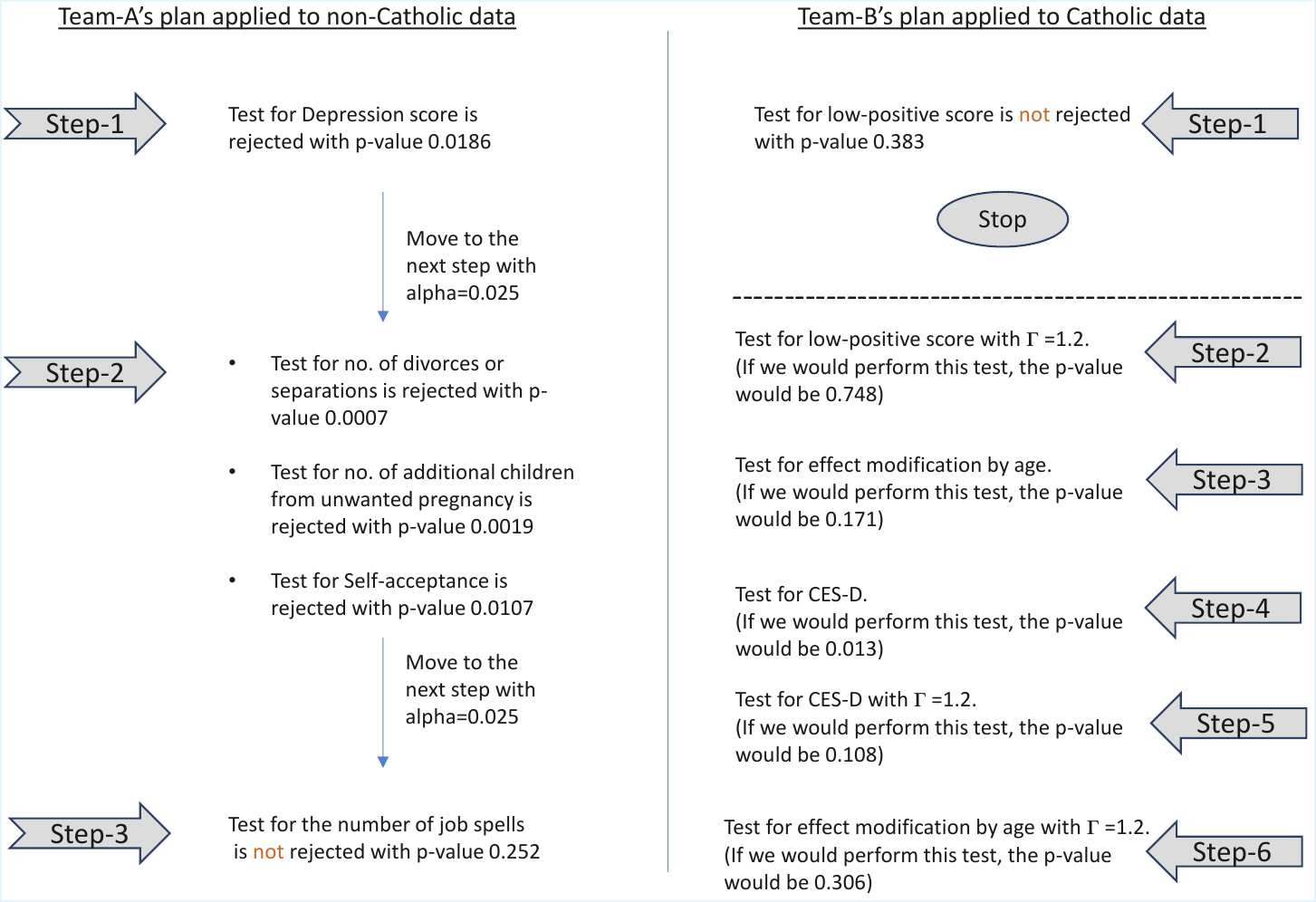}
    \caption{Results of the two team cross-screening: depression score, self-acceptance, number of divorces or separations and additional number of children from unwanted pregnancy are rejected global nulls. No replicable outcomes are detected.}
    \label{fig:twotm}
\end{figure}
As depicted in Figure \ref{fig:twotm}, while applying team A's plan (see Figure \ref{fig:test_seq}) on the non-Catholics data, the test for depression score was rejected in the first step, and the testing procedure moved forward to the second step. In the second step, the test for the number of divorces or separations, the test for number of additional children from unwanted pregnancies, and the test for self-acceptance were rejected. Thus the process moved further to the third step. However, in this step, the test for the number of job spells was not rejected. On the other hand, while applying team B's plan on the Catholics data, the test for low-positive affect score was not rejected at the first step, and the testing procedure ended there.

Our results reveal new insights into the lasting effects of unwanted pregnancies on mothers. The only prior long-term study we know of \citep{herd2016implications} found adverse effects on depression. We not only confirmed this, but also found significantly lower self-acceptance among mothers with unwanted pregnancies, reinforcing short-term evidence from \citet{biggs2014does}. We further observed higher divorce rates and a greater likelihood of subsequent unwanted pregnancies, underscoring important implications for policy and awareness programs.


The above effects were found only using team A’s plan on non-Catholic data; the other team’s plan was terminated in the first step (Figure \ref{fig:twotm}). Hence we did not produce any replicable findings. It is not clear why the effect of unwanted pregnancy on the low-positive affect subscale was only significant among non-Catholics and not among Catholics. This could be explored in future research. 
\section{Comparison of Results to Existing Approaches}
\label{auto_results}
We compared our findings with the ones obtained by some other relevant approaches described in \S~C of SM-1. We chose to compare our findings with those obtained by automated cross-screening \citep{cross-screening} and Holm on the full data \citep{holm1979simple} because these were the most powerful competing approaches for replicability analysis and global null testing respectively in the simulations in SM-1. 

\begin{figure}[h!]
    \centering
    \includegraphics[scale=0.6]{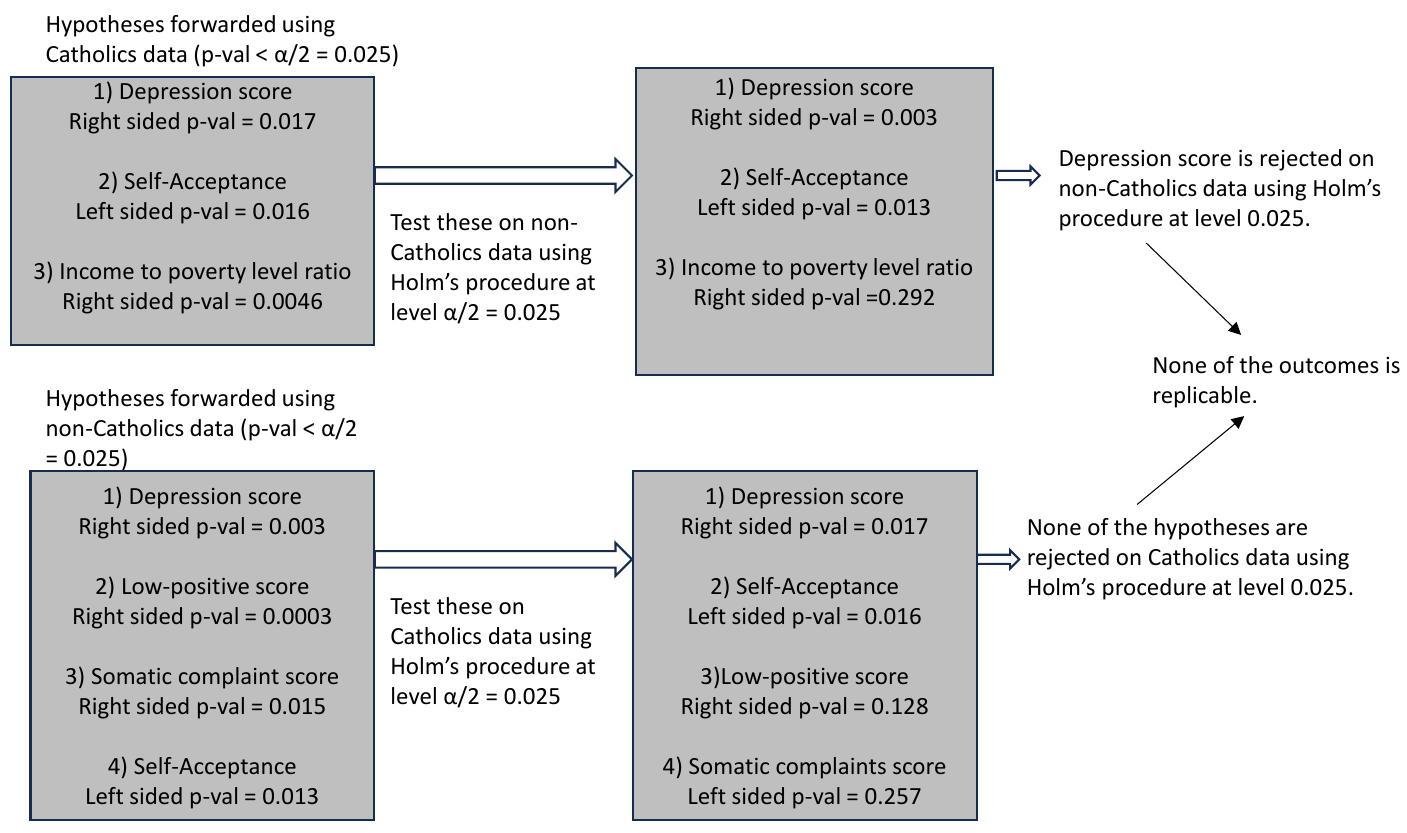}
    \caption{Results of automated cross-screening. Right-sided (or, left-sided) p-value corresponds to the alternative that the outcome is higher (or, lower) for the women with unwanted pregnancies compared to their controls.}
    \label{fig:auto}
\end{figure}

Figure \ref{fig:auto} depicts the findings of applying automated cross-screening to our data. At the first stage of automated analyses in each part of the data to choose hypotheses to test in the other part of the data, the tests for depression score, self-acceptance and income to poverty level ratio were rejected on the Catholic data and the tests for depression score, self-acceptance, low-positive score and somatic complaint score were rejected on the non-Catholic data. When these hypotheses were tested on the other part of the data using Holm's procedure at level 0.025, depression score had a significant effect in the non-Catholic data and there were no significant effects in the Catholic data. Thus, there were no replicable findings.  

For testing of global nulls, Holm on the full data found that depression score and low-positive subscale score were significantly higher among women with unwanted pregnancies.  

Automated cross-screening and Holm-full data are prespecified approaches whose analysis plan must be prespecified before looking at the data while two team cross-screening allows the teams to explore their parts of the data to prepare an analysis plan for the other part of the data. This exploration creates the opportunity to delve into new unanticipated hypotheses in addition to a prefixed set of outcomes. For our data, two team cross-screening found effects on several outcomes that were not considered in the prespecified approaches, namely self-acceptance, number of divorces or separations, and number of additional children from unwanted pregnancies (see \S~\ref{results}).  

Another advantage of the two team cross-screening over prespecified approaches is that the EDA can inform us what hypotheses are meaningful and we can focus our error budget $\alpha$ on those hypotheses.  In the left panel of Figure \ref{fig:auto}, income to poverty level ratio appeared influential among Catholics, so the automated approach proposed testing this on non-Catholics. In contrast, Team A found that the effect was driven solely by pensions, annuities, and survivor’s benefits—an uninterpretable pattern—and chose not to pursue it (\S~\ref{plan_a}). Thus, by carefully examining the data before finalizing the analysis plan, two team cross-screening avoided the use of $\alpha$ on an uninformative hypothesis.


\section{Two Team Cross-Screening for Effect Sizes}
\label{esize}

We have focused on using two team cross-screening for hypothesis testing. However, it can also support the construction of confidence intervals (CIs) for effect sizes by helping to identify outcomes for which CIs are of particular interest. 

\textcolor{blue}{
In a typical study examining multiple outcomes, the analyst is interested in the effect size only for a selected subset of outcomes that seem interesting. While 95\% CIs guarantee that the expected non-coverage is only 5\% across all the parameters, the non-coverage may be much higher among the selected. In the extreme case that all parameters are null, and the analyst selects the parameters with 95\% CIs that do not cover the null value, the expected non-coverage is one. Therefore, it is necessary to adjust for the selection effect in order to have a meaningful error guarantee such as simultaneous non-coverage or false coverage rate (FCR) at most $\alpha$ \citep{benjamini2005false, Frostig2024}. The adjustment for selection necessarily entails making the CIs wider, and the resulting adjusted CIs may be too wide to be of interest to the analyst. 
}

\textcolor{blue}{
Using the two team approach, the inflation in CI width may be avoided by letting each team analyze its data to determine the outcomes for which to report a CI on the other part of the data. 
The selection criterion the teams use might be different than for testing hypotheses.  Any outcome that seems to have signal and is of scientific interest might be worth including when forming CIs even if its one-sided p-value is  greater than 0.025. 
If a CI is constructed for a parameter selected by either Team A or Team B at level $1-\alpha/2$, then  the expected proportion of constructed confidence intervals that do not cover the truth among all constructed CIs (i.e., the FCR) is at most $\alpha$. This follows since the data used for the selection of parameters are independent of the data used to construct the confidence intervals, see \S~E of SM-1 for the simple proof. 
}

\textcolor{blue}{
If control of a stricter  error rate is desired, one option is the overall simultaneous non-coverage. The following  approach would guarantee 95\% overall simultaneous coverage (and more strongly at most 2.5\% chance of not covering at least one outcome in each subgroup): for the $n_1$ outcomes specified by Team A for CI construction on Team B’s data, form $(1-0.025/n_1)$ CIs (e.g., CIs based on the $(1-0.0125/n_1)$ quantiles of a pivotal statistic) on Team B's data, and for the $n_2$ outcomes specified by Team B for CI construction on Team A’s data, form $(1-0.025/n_2)$ CIs on Team A’s data. A proof is in \S~E of  SM-1.
}
\section{Discussion}
\label{disc}
We proposed a novel two team cross-screening approach that enabled us to perform exploratory data analysis, confirmatory data analysis and replication in the same observational study. The chance of finding replicable outcome could be improved in future studies by having more pre-analysis discussion. Team A was more familiar with the WLS database than team B and considered more additional outcomes beyond those decided during the pre-analysis discussion; a lengthier pre-analysis discussion might have led to more uniform choices. Both teams chose their tests for the other subgroup based on minimizing p-values on the subgroup they analyzed. It might have been better to use more adaptive tests (e.g., \citealt{rosenbaum2020combining}), recognizing the role of chance in making one test look best on a subgroup as well as the fact that effect sizes might differ on the other subgroup.

\textcolor{blue}{A reviewer raised the question of whether unwanted pregnancy for Catholic and non-Catholic women represents the same exposure. In the era after {\it{Roe v. Wade}} and before {\it{Dobbs v. Jackson}} where abortion was legal everywhere in the U.S., Catholic and non-Catholic women may have interpreted ``unwanted pregnancy" differently. Catholic women who carried unintended pregnancies to term might disproportionately represent cases that were firmly unwanted, while non-Catholic women might disproportionately represent cases of ambivalence (since more strongly unwanted pregnancies could end in abortion). However, in the historical context of our data {\textendash} when abortion was illegal for all women in Wisconsin {\textendash} the structural reality of the exposure was much more homogeneous: women carried to term pregnancies that they did not intend, regardless of religious affiliation. 
}

There can be more than one feature to split the data on. In our study, in addition to Catholics and non-Catholics having unwanted pregnancies for relatively different reasons, women who graduated from college and women who did not might have unwanted pregnancies for relatively different reasons. One option for making use of two ways to split the data is the following. Team A can explore Catholics with college education and non-Catholics without college education and specify two different analysis plans, one for Catholics without college education and another for non-Catholics with college education, each controlling for $0.05/4$ FWER. Team B can explore Catholics without college education, and non-Catholics with college education and specify two different analysis plans – one for Catholics with college education and another for non-Catholics without college education – each controlling for $0.05/4$ FWER. The FWER is $0.05$ for a finding in the confirmatory analysis in any of the four subgroups. Another option for making use of two ways to split the data is four team cross-screening, where each of four teams can design a $0.05/4$ analysis plan for one subgroup using the data from the remaining three subgroups. Like the first option, the FWER is $0.05$ for a finding in the confirmatory analysis in any of the four subgroups. These approaches of considering multiple splits allow one to consider whether there is replicability across multiple splits. 

\textcolor{blue}{We use only two teams in this work. Why not consider three or more?  A clear disadvantage is the added logistical complexity. On the other hand, a potential  advantage is the possibility of  stronger replicability claims and  more creative exploration (as suggested by a reviewer). However, the best way to organize analyses with more than two teams remains an open question. 
Consider, for example, three teams. One option is to assign each team a single subgroup for exploration, to guide the analysis of the remaining subgroups. This leaves most of the data available for inference, but requires methodological advances in how to unify multiple analysis proposals for the same subgroup. Our simulations (see 
\S~F of SM-1) indicate that a naive unification approach offers no meaningful power improvement over the two-team setting. A second option is to give each team access to all but one subgroup, tasking them with designing the analysis for the subgroup they did not see. Inference is straightforward because each subgroup has exactly one proposed analysis, but most of the data is sacrificed for exploration. Our simulations suggest that this approach may be less powerful unless exploration is effective enough to yield substantially improved testing strategies. It is essential that such strategies take into account that the data available for testing is much smaller than that for exploration. How to achieve this effectively is a direction for future research. 
}



While having many advantages, two team cross-screening has some limitations. Obviously, two team cross-screening is not feasible when there is only one person working on the project. Moreover, if there is effect modification by the covariate used to split the data, then using one subgroup to design the analysis of the other subgroup may not be effective since learning the effect pattern in one subgroup may tell us little or nothing about the effect pattern in the other subgroup. If the researchers worry about effect modification, the following variant of two team cross-screening may be considered instead: split the data randomly into three parts $\textendash$ say 20\%, 20\% and 60\% $\textendash$ and then team A will use the first 20\% to design the analysis for the remaining 80\% and team B will use the second 20\% to design the analysis for the other 80\% (first 20\% and last 60\%). Team A could propose tests for both subgroups and thus have a chance to achieve replicability; likewise, team B could propose tests for both subgroups. Like in the two team cross-screening approach discussed in this paper, this variant also uses all the data for inference, but here each team uses less data for planning but more for inference. 

This paper has focused on using two team cross-screening for exploration, confirmation and replication in observational studies, but the approach may be useful for other types of studies.  For randomized trials, replication in multiple trials (which includes replication across two or more sites in a single study) can enhance external validity \citep{barnow2020conducting}. In fact, in many situations, the U.S. Food and Drug Administration (FDA) requires two adequate and well controlled trials to establish effectiveness of a drug because ``two positive trials differences in design and conduct may be more persuasive as unrecognized design flaws or biases in study conduct will be less likely to impact the outcomes [and the] consistency of results across two trials also greatly reduces the possibility that a biased, chance, site-specific or fraudulent result will lead to an erroneous conclusion” \citep{FDA2019SubstantialEvidence}. To ensure rigorous confirmatory analyses, the FDA currently requires that trials adhere to strict protocols for data analysis that are written before the trial is conducted, limiting the scope for EDA.  The use of two team cross-screening would enable EDA to be used while still providing rigorous error control for confirmation and replication.  Another example in which two team cross-screening could be useful is in testing for unobserved heterogeneity. \cite{patton2023testing} proposed a split sample procedure in which the data is split into two parts, e.g., two time periods, and one part of the data is used to choose clusters and the other part is used to test whether the clusters have different means.  A confirmatory analysis is only done for the second part of the data, so there is no opportunity for replication. Two team cross-screening would enable EDA to still be used in choosing the clusters but also allow for confirmatory analysis in both parts of the data, thus providing the possibility of replication. In genome-wide association studies, it is standard to use a discovery cohort to identify candidate SNPs associated with the trait and a replication (or validation) cohort to test only those candidates and confirm that the association holds \citep{Marigorta2018}. With two-team cross-screening, exploratory data analysis can be incorporated to optimize design decisions for testing SNP–trait associations, while also enabling the reporting of replicability findings with control over false replicability claims.



\bigskip
\begin{center}
{\large\bf Supplementary Material}
\end{center}
SM-1 provides detailed description on the outcomes and simulation studies. SM-2 contains the protocol document. The R codes and master data file are provided in this \href{https://github.com/anonymousSR67/Two_team_Cross_Screening/tree/main}{GitHub} and \href{https://www.dropbox.com/scl/fo/oalt1prlim73feoq74bgb/AOiwvwCmLYr4PTafnMXxAUg?rlkey=oqhaykh7nn5d82o74yxo9gxt7&st=l5ljv3lt&dl=0}{this link} respectively.

\bibliographystyle{apalike}

\bibliography{JASA-template}

\end{document}